\documentclass{article}
\usepackage{graphicx}
\usepackage{graphics}

\usepackage[margin=1in]{geometry}

\usepackage{bbold}
\usepackage{xparse}
\usepackage{amsmath}
\usepackage{xcolor}

\makeatletter \newcommand{\IfNoValueOrEmptyTF}[3] {\IfNoValueTF{#1}{#2}{\def\@tempa{#1}\ifx\@tempa\@empty#2\else#3\fi}} \makeatother

\DeclareDocumentCommand \rs {o} {%
  \IfNoValueTF{#1}{\mathbb{R}}{\mathbb{R}^{#1}}%
}
\DeclareDocumentCommand \cs {o} {%
  \IfNoValueTF{#1}{\mathbb{C}}{\mathbb{C}^{#1}}%
}

\newcommand{\ve}[1]{\mathbf{#1}}

\DeclareDocumentCommand \vet {mo}{%
\mathbf{#1}\IfNoValueOrEmptyTF{#2}{(t)}{(#2)}
}
\DeclareDocumentCommand \dvet {mo}{%
\dot{\mathbf{#1}}\IfNoValueOrEmptyTF{#2}{(t)}{(#2)}
}

\newcommand{\mat}[2]{\left [ \begin{array}{#1} #2 \end{array}\right ]}

\usepackage{tikz}
\usetikzlibrary{calc}
\usepackage{stmaryrd}

\usepackage{blkarray}

\usepackage{graphicx}
\usepackage{enumitem}

\newlist{inlinelist}{enumerate*}{1}
\setlist*[inlinelist,1]{%
  label=(\roman*),
}

\usepackage{array}
\newlength{\mycolwidth}
\newlength{\mycolwidthh}
\usepackage{array}
\newcolumntype{Z}{>{\centering\arraybackslash$}p{\mycolwidth}<{$}}
\newcolumntype{W}{>{\centering\arraybackslash$}p{\mycolwidthh}<{$}}

\begin{document}

\title{Performance evaluation of gust load alleviation systems for flexible aircraft via optimal control
}

\author{Pierre Vuillemin$^\dagger$, David Quero Martin$^\ddagger$ \& Charles Poussot-Vassal$^\dagger$\\
\small
$^\dagger$ONERA / DTIS, Universit\'e de Toulouse, F-31055 Toulouse, France \\
\small \texttt{pierre.vuillemin@onera.fr}, \texttt{charles.poussot-vassal@onera.fr} \\
\small $^\ddagger$DLR (German Aerospace Center), Institute of Aeroelasticity, Aeroelastic Simulation, G\"ottingen, Germany \\
\small \texttt{David.QueroMartin@dlr.de}
}

\date{}

\maketitle

\begin{abstract}
The dynamical response of an aircraft subject to gust perturbations is a key element in a preliminary design phase. In particular, the loads induced by gusts along the wing should not exceed some limit values and should even ideally be decreased. Active control is one lever to address this problem. However, evaluating the benefit that active control may bring considering some actuators characteristics or some delay in the loop is a difficult task, especially in the early design phase. This problem is addressed in this paper with an open-loop optimal control framework and more specifically with a direct transcription method resulting in a linear optimisation problem. The approach is illustrated on a realistic aeroelastic aircraft model built with a coupled fluid-structure solver which order is reduced to decrease the number of optimisation variables. 
\end{abstract}

\section{Introduction}

Designing or optimising the structure and geometry of an aircraft is a challenging task for aeronautical engineers as they have to cope with several, often contradictory, objectives and constraints. The latter are often related to the specifications stated by flight authorities that must be met before any aircraft is allowed to fly. One of these specifications concerns the behaviour of the aircraft in response to a predefined set of gust disturbances.

Such perturbations can indeed lead to an important stress on the aircraft structure and especially the wings which may endure important loads. The latter need to be monitored to ensure that they do not exceed some limit values to guarantee the structural integrity of the aircraft in all flight conditions. This clearance constraint is contradictory to the efforts made to lighten the aircraft structure in order to meet modern aircraft consumption and overall traffic footprint impact reduction objectives. Indeed, lighter aircraft tend to be more flexible and thus more sensitive to gust disturbances. In that context, active control plays a crucial role to alleviate the loads on the wing when gusts are encountered. Therefore, when modern flexible aircraft are designed or optimised, the possible gain brought by active control must be evaluated very early in the design process. 

\paragraph{Loads alleviation problem \& contributions. }
Let us consider a linear time invariant (LTI) model in standard form representing the linearised dynamical behaviour of an aeroelastic aircraft around some flight point,
\begin{equation}
  \begin{array}{rcl}
    \dvet{x}&=& A   \vet{x} + B_1    \vet{w} + B_2    \vet{u} \\
    \vet{z} &=& C_1 \vet{x} + D_{11} \vet{w} + D_{12} \vet{u} \\
    \vet{y} &=& C_2 \vet{x} + D_{21} \vet{w} + D_{22} \vet{u}
  \end{array}
  \label{eq:sys}
\end{equation}
where $\vet{x} \in \rs[n]$ is the state, $\vet{u} \in \rs[n_u]$ the control input, $\vet{w} \in \rs[n_w]$ the exogenous input, $\vet{y} \in \rs[n_y]$ the measurement output and $\vet{z}\in\rs[n_z]$ the performance output. 

In the context of gust load alleviation,  the control input $\ve{u}$ is usually composed of the elevators, inner and outer ailerons but could be completed with additional inputs within the context of novel aircraft design such as actuators for flow control, etc. Depending on the cases, the performance output $\ve{z}$ may represent the bending moments or the vertical load factor along the wing. The measurements can include usual flight dynamics quantities (angle of attack, etc.), accelerometers along the fuselage or more unconventional sensors in the context of novel aircraft design.

To design an active control law, the quantity of interest is the envelope formed by the worst bending moments obtained with time-domain simulations when the model \eqref{eq:sys} is excited with a set of specific gust profiles. The latter are the so-called $1-cosine$ gust profiles as detailed for instance by \cite{Tang:1996} and which are used for certification. In practice, the active control law is generally sought as a feedback law linking the  measured output $\ve{y}$ to the control input $\ve{u}$ and designed to decrease the worst-case envelope mentioned above. This can for instance be achieved by minimising the $\mathcal{H}_\infty$-norm of the channel from $\ve{w}$ to $\ve{z}$ with adequate filters as it has been done for instance by the authors in the industrial application (\cite{poussot:2020:gla}). However, feedback control design can be a difficult task requiring several trials to adjust the tuning parameters in a satisfactory way. It becomes even more difficult when actuators saturations of delays have to be considered or when the state dimension $n$ of the model \eqref{eq:sys} is large.

For these reasons and in an early design perspective, a direct open-loop optimal control framework (\cite{betts:practical:2010}) is used instead. Its objective is to help characterise the controlled performances of an aircraft with respect to gust load alleviation. This is done considering the operational limitations of the actuators, such as input magnitude and rate saturations, and the presence of eventual delays between the moment the gust reaches the aircraft and the control starts acting. Such framework gives interesting insights concerning control efficiency of the aircraft and the associated limiting factors. It could for instance be integrated in an automated design process or help to assess the potential benefits offered by modern technologies such as LIDAR as explored in the work of \cite{khalil:2021:gla}. Still, the dimension of the model \eqref{eq:sys} remains an issue as it directly impacts the number of variables in the transcribed optimisation problem. This is addressed through model reduction techniques (see the book by \cite{antoulas:approximation:2005} for an overview) which resulting error can be bounded and incorporated within the reduced optimisation problem at the cost of an increased conservatism.

The overall optimisation framework and objectives considered in this article are similar to those of the work by \cite{haghighat:model:2012} and \cite{giesseler:mpc:2012} but several distinctions can be made: since it is sought as a design tool rather than an embedded solution, no receding horizon scheme is considered and additional constraints can be added to enrich the problem. In particular, robustness elements are considered to account for the variability of the gust profiles and potentially the error induced by the reduction step. On a more technical side, a $l_\infty$ objective function is considered instead of a $l_2$ objective thus resulting in a linear optimisation problem. 
While the approach could in theory be applied online in a receding horizon scheme, several requirements should be met for a real-world scenario such as an efficient state observer, bounded worst execution time, proof of stability, guarantee on the convergence of the optimisation algorithm. These questions are out of the scope of this article.

\paragraph{Paper structure. } 
In Section \ref{sec:optimal_control}, the gust load alleviation problem is formulated and recast as a linear optimisation one through a direct transcription approach. Then in Section \ref{sec:appli}, the optimal control problem is solved for a realistic aeroelastic aircraft model to highlight the impact of the actuators and the delays on the performances for gust load alleviation. Finally, conclusive remarks are given in Section \ref{sec:ccl} together with some perspectives concerning future works.

\section{Optimal gust load alleviation problem}
\label{sec:optimal_control}

In this section, the gust load alleviation problem is transcribed into an optimal control problem that is both versatile to allow for relevant constraints formulation and also  tractable for an efficient practical resolution. In particular, in section \ref{ssec:general_problem}, the gust load alleviation problem is stated more formally in continuous-time. The finite dimensional optimisation problem obtained after discretisation is then presented in section \ref{ssec:transcription}.  The latter is further recast as an equivalent linear optimisation problem in section \ref{ssec:equivalent_lp}. Finally, the means to decrease the number of variables are described in section \ref{ssec:nvars}.

\subsection{General problem formulation}
\label{ssec:general_problem}

As already mentioned in the introduction, the overall objective of gust load alleviation control is to reduce the loads endured by the wing when encountering some gust. To translate this objective more formally, let us consider the linearised aeroelastic model \eqref{eq:sys}. The latter is assumed to be stable. Due to the open-loop nature of the approach, the measurement output $\ve{y}$ is not exploited in the sequel. The considered formulation of the constraints allows to consider a wide variety of control input and performances output. However, for sake of simplicity and to state clearly the ideas that led to this formulation, the specific configuration that inspired it is recalled. It is similar to the one described by \cite{poussot:2020:gla} and is common in gust load alleviation control: the control input consists in inner and outer ailerons and the elevator, the gust input consists in the vertical wing speed at the nose of the aircraft and the performance output consists in several bending moments along the wing.

The linear dynamical model \eqref{eq:sys} is fed on the time interval $\mathcal{I}$ with a control input $\ve{u} \in \mathcal{U}\subseteq \mathcal{L}_2$ and a perturbation $\ve{w}$ taken from a finite set $ \mathcal{W} = \lbrace \ve{w}^{(1)},\ldots,\ve{w}^{(n_p)}  \rbrace \subseteq  \mathcal{L}_2$ of gust profiles. Let us denote the finite set of resulting performance outputs by $\mathcal{Z}(\ve{u}, \mathcal{W}) =\lbrace \ve{z}^{(1)},\ldots,\ve{z}^{(n_p)} \rbrace \subseteq \Xi $. The objective is then to minimise the worst performance output considering all the gust perturbations. More formally, one wants to find the optimal control signal $\ve{u}^\star$ that solves

\begin{equation}
\min_{\ve{u} \in \mathcal{U}} \mathcal{J}_\ve{z} = \max_{ \ve{z} \in \mathcal{Z}(\ve{u}, \mathcal{W})} \Vert  W_\ve{z}  \ve{z} \Vert_{\infty}
\label{eq:init_pb}
\end{equation}
where for all $\ve{w} \in \mathcal{W}$, the corresponding $\ve{z}$ satisfies \eqref{eq:sys} and $W_\ve{z} \in \mathbb{R}^{n_z \times n_z}$ is some weighting matrix that can for instance be used to normalise the various elements of $\ve{z}$.

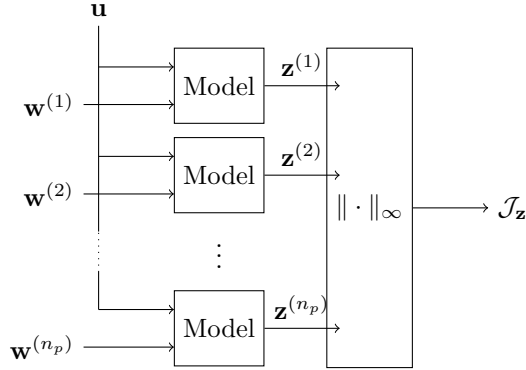
\begin{figure}
  \centering
  \begin{tikzpicture}

   \node (M1)[anchor = north,draw,minimum height= 1cm] at (0,0) {Model};

   \node (M2) [draw,anchor = north, below = 5pt,minimum height= 1cm] at (M1.south) {Model};
   \node (M3) [anchor = north, below = 2pt] at (M2.south) {$\vdots$};
   \node (M4) [draw,anchor = north, below = 5pt,minimum height= 1cm] at (M3.south) {Model};

   \node (u) at ($(M1.west) + (-1,1)$) {$\ve{u}$};
   \coordinate (u1) at ($(M1.west) + (0,0.25cm)$);
   \coordinate (u2) at ($(M2.west) + (0,0.25cm)$);
   \coordinate (u4) at ($(M4.west) + (0,0.25cm)$);
   \coordinate (u1i) at ($(u1) + (-1,0)$);
   \coordinate (u2i) at ($(u2) + (-1,0)$);
   \coordinate (u4i) at ($(u4) + (-1,0)$);

   \draw [->] (u1i) -- (u1);
   \draw [->] (u2i) -- (u2);
   \draw [->] (u4i) -- (u4);

   \draw (u.south) -- ($(u2i)+(0,-1)$);
   \draw (u4i) -- ($(u4i)+(0,0.5)$);
   \draw [dotted]($(u4i)+(0,0.5)$) -- ($(u2i)+(0,-1)$);

   \coordinate (w1) at ($(M1.west) + (0,-0.25cm)$);
   \coordinate (w2) at ($(M2.west) + (0,-0.25cm)$);
   \coordinate (w4) at ($(M4.west) + (0,-0.25cm)$);

   \node (w1i) [anchor = east] at ($(w1) + (-1.2,0)$) {$\ve{w}^{(1)}$};
   \node (w2i) [anchor = east] at ($(w2) + (-1.2,0)$) {$\ve{w}^{(2)}$};
   \node (w4i) [anchor = east]  at ($(w4) + (-1.2,0)$) {$\ve{w}^{(n_p)}$};

   \draw [->] (w1i.east) -- (w1);
   \draw [->] (w2i.east) -- (w2);
   \draw [->] (w4i.east) -- (w4);


   \draw [->] (M1.east) -- ($(M1.east) + (1,0)$);
   \draw [->] (M2.east) -- ($(M2.east) + (1,0)$);
   \draw [->] (M4.east) -- ($(M4.east) + (1,0)$);

   \node [anchor = south] at ($(M1.east) + (0.5,0)$) {$\ve{z}^{(1)}$};
   \node [anchor = south] at ($(M2.east) + (0.5,0)$) {$\ve{z}^{(2)}$};
   \node [anchor = south] at ($(M4.east) + (0.5,0)$) {$\ve{z}^{(n_p)}$};

   \node (m) [draw,anchor = north,minimum height = 4.25cm] at (2,0) {$\Vert \cdot \Vert_{\infty}$};

   \node (J) [anchor = west] at ($(m.east) + (1,0)$) {$\mathcal{J}_\ve{z}$};
   \draw [->] (m.east) -- (J.west);
\end{tikzpicture}
  \caption{Schematic interpretation of the criterion $\mathcal{J}_z$. The control input $\ve{u}$ feeds in parallel several instances of the same model \eqref{eq:sys} which are disturbed by different gusts $\ve{w}^{(i)}$. The resulting worst-case performance forms the objective function $\mathcal{J}_{\ve{z}}$. }
  \label{fig:opti_pb}
\end{figure}

Problem \eqref{eq:init_pb} is represented schematically in figure \ref{fig:opti_pb}. The problem can be considered as robust in the sense that \emph{only one single} command input signal is sought to compensate for the effect of a (discrete) set of disturbances. It enables to account for some uncertainty associated with the gusts and echos multi-models design in robust control or more generally scenario optimisation. The way this uncertainty can be exploited is illustrated further in section \ref{sec:appli}.

The remainder of this section specifies the problem more in detail through the definition of the constraints represented by the control input set $\mathcal{U}$, the performance set $\Xi$ and a complementary objective in the cost. The specific robustness constraints that can be incorporated to the problem to compensate for the use of a reduced-order model instead of the large one are described later in section \ref{ssec:nvars}.

\paragraph{Input constraints. }The set of admissible control input $\mathcal{U}$ can be used to translate the operational limitations of the actuators. The most common constraints associated with the actuators available for gust load control on an aircraft are the magnitude and rate saturations of control surfaces. The input set $\mathcal{U}$ is thus defined as,
\begin{equation}
    \mathcal{U} = 
    \left \lbrace 
    \ve{u}:t\to\ve{u}(t)
    \big/ \forall t \in \mathcal{I},\, |\ve{u}(t)| \leq \overline{\ve{u}}\text{ and } |\dot{\ve{u}}(t)| \leq \overline{\overline{\ve{u}}}
    \right \rbrace.
\end{equation}
where $\overline{\ve{u}}>0$ and $\overline{\overline{\ve{u}}}>0$ belongs to $\mathbb{R}^{n_u}$ and are the values of the saturations which could as well be made asymmetric. Another simple refinement of this set consists in allowing the saturation $\overline{\ve{u}}$ to be a function of the time. For instance, with a step, the control input can be allowed to be non-null before the perturbation starts to simulate a pre-emptive action, or conversely to simulate a delay. This is exploited in section \ref{sec:appli}.

\paragraph{Performance constraints. }Since the objective $\mathcal{J}_\ve{z}$ is solely focused on the worst case, a control may decrease an element of $\ve{z}$ (the worst) while increasing another one. To avoid such an issue and to enforce that any control input must at least preserve the natural behaviour of the aircraft, one can adjust the set of admissible performance output $\Xi$. In particular, let us consider $\mathcal{Z}(0,  \mathcal{W})$, the set of performance output obtained with no control action and define $\overline{\ve{z}} \in \mathbb{R}^{n_z}$ and $\underline{\ve{z}}\in\mathbb{R}^{n_z}$, for $i=1,\ldots,n_z$, as,
\begin{equation}
    \ve{e}_i^T\overline{\ve{z}} = \max_{\ve{z} \in \mathcal{Z}(0,  \mathcal{W})} \Vert \ve{e}_i^T\ve{z} \Vert_\infty\quad\text{and}\quad \ve{e}_i^T\underline{\ve{z}} = -\max_{\ve{z} \in \mathcal{Z}(0,  \mathcal{W})} \Vert - \ve{e}_i^T \ve{z}  \Vert_\infty,
    \label{eq:wcp}
\end{equation}
 where $\ve{e}_i \in \mathbb{R}^{n_z}$ is the $i$-th canonical vector. Equation \eqref{eq:wcp} represents the element-wise worst positive and negative performance values, respectively. Then the set $\Xi$ of admissible performance output should be defined as, 
\begin{equation}
    \Xi = \left \lbrace \ve{z}:t\to \ve{z}(t) \big / \forall t \in \mathcal{I}, \underline{\ve{z}} \leq \ve{z}(t) \leq \overline{\ve{z}}
    \right \rbrace.
    \label{eq:perf_cst}
\end{equation}
By adjusting the bounds $\overline{\ve{z}}$ and $\underline{\ve{z}}$, the set $\Xi$ could be used to try to impose minimal performances. The resulting problem might not be feasible anymore thus indicating that the prescribed level of performance cannot be reached with the considered configuration.

\paragraph{Complementary energy objective. }Similarly, looking only at the worst case may lead to an infinite number of control signals achieving the same performances. To specify the solution further, one could complete $\mathcal{J}_\ve{z}$ with a secondary objective $\mathcal{J}_\ve{u}$ aimed at promoting control signals with the lowest energy. This approach can be referred to as scalarisation of a multi-objective function. While the $l_2$-norm is the natural choice, the transcribed optimisation problem could not be turned into an equivalent linear problem in section \ref{ssec:equivalent_lp}. Therefore, one can instead consider the $l_1$-norm so that  $\mathcal{J}_\ve{u} = \epsilon \Vert \ve{u} \Vert_1$ where $\epsilon >0$ is a tuning parameter to control the importance of this secondary objective with respect to the primary one $\mathcal{J}_\ve{z}$. 

\paragraph{Feasibility. }At this point, one can observe that the natural (uncontrolled) behaviour of the aircraft, i.e. for $\ve{u} = 0$, is always feasible with respect to the constraints of Problem \eqref{eq:init_pb}. Indeed, it is clear that $0 \in \mathcal{U}$ and by definition, $\forall t \in \mathcal{I}$, $\forall \ve{z} \in \mathcal{Z}(0,  \mathcal{W})$, $\underline{\ve{z}} \leq \ve{z}(t) \leq \overline{\ve{z}}$.

\subsection{Transcription of the optimal control problem}
\label{ssec:transcription}

To address problem \eqref{eq:init_pb} in practice, a direct approach is considered, see for instance the book by \cite{betts:practical:2010} for an overview of these techniques. More specifically, by discretising the time, the optimal control problem \eqref{eq:init_pb} can be approximated by a standard finite dimensional optimisation problem.

For that purpose, the time interval $\mathcal{I}$ is divided into $N-1$ equals sub-intervals $[t_k, t_{k+1}]$ corresponding to a constant sampling time $h>0$. The sought control signal $\ve{u}$ is approximated by a piecewise constant function over $\mathcal{I}$ such that
\begin{equation}
    \forall k \in \llbracket 0,N-1\rrbracket ,\, \forall t \in [t_k,t_{k+1}[,\,\ve{u}(t) = \ve{u}_k
\end{equation}
and is naturally parametrised by the discrete set of $N$ values $\left \{ \ve{u}_k\right \}_{k=0}^{N-1}$. The linear dynamical model \eqref{eq:sys} is discretised accordingly with a zero-order hold method such that, $\forall k \in \mathbb{N}$,
\begin{equation}
H:\left \lbrace 
   \begin{array}{rcl}
    \ve{x}_{k+1}&=& \mathcal{A}   \ve{x}_k + \mathcal{B}_1    \ve{w}_k + \mathcal{B}_2    \ve{u}_k \\
    \ve{z}_k &=& \mathcal{C}_1 \ve{x}_k + \mathcal{D}_{11} \ve{w}_k + \mathcal{D}_{12} \ve{u}_k \\
  \end{array}
  \right .
  \label{eq:dsys}
\end{equation}
where $\ve{x}_k$, $\ve{z}_k$ and $\ve{w}_k$ are the values of the corresponding continuous signals at the sampling time $t_k$. 

Let $\mathcal{K}$ denote the set of time indices from $0$ to $N-1$ and $\mathcal{P}$ the set of $n_p= card(\mathcal{W})$ disturbance indices. Each disturbance profile in $\mathcal{W}$ leads to a different trajectory for the model \eqref{eq:dsys}, let us use a superscript to distinguish them, so that for $p \in \mathcal{P}$, the corresponding gust profile and associated state and performance at time $k$ are denoted $\ve{w}_k^{(p)}$, $\ve{x}_k^{(p)}$ and $\ve{z}_k^{(p)}$, respectively. With these notations, the gust load alleviation problem \eqref{eq:init_pb} then becomes,
\begin{equation}
  \begin{array}{rc}
     \displaystyle \min_{\left \{ \ve{u}_k\right \}_{k\in \mathcal{K}}} & \displaystyle\max_{k\in \mathcal{K}} \max_{p\in \mathcal{P}} \Vert W_\ve{z} \ve{z}_k^{(p)} \Vert_{\infty} + \epsilon \sum_{k\in \mathcal{K}} \Vert \ve{u}_k \Vert_1 
     \\
     s.t.\, \forall k \in \mathcal{K},\,\forall p\in \mathcal{P}&
     
     \begin{array}[t]{rcl}
        \ve{x}_{k+1}^{(p)}&=&\mathcal{A} \ve{x}_k^{(p)} + \mathcal{B}_{1} \ve{w}_k^{(p)}+ \mathcal{B}_2\ve{u}_k\\
        \ve{z}_{k}^{(p)}&=&\mathcal{C}_1 \ve{x}_k^{(p)} + \mathcal{D}_{11} \ve{w}_k^{(p)}+ \mathcal{D}_{12}\ve{u}_k\\
        \left | \ve{u}_k \right |&\leq& \overline{\ve{u}}\\
        \left |\ve{u}_{k+1} - \ve{u}_k\right |&\leq&h\overline{\overline{\ve{u}}}\\
        \ve{z}_k^{(p)}&\leq& \overline{\ve{z}}\\
        \ve{z}_k^{(p)}&\geq& \underline{\ve{z}}\\
        \ve{x}_0^{(p)} &=&\ve{x}(0)
     \end{array}
  \end{array}
  \label{eq:mainPb}
\end{equation}

As its objective and constraints are convex, problem \eqref{eq:mainPb} is convex. Indeed, by exploiting the rules stated in \cite[chap.3]{boyd:convex:2004}, one can see that $\Vert W_\ve{z} \ve{z}_k^{(p)} \Vert_{\infty}$ is convex as the composition of the convex function $\Vert \cdot \Vert_\infty$ with an affine function and since the point-wise maximum of convex functions is also convex, then the function $\max_{p\in \mathcal{P}} \Vert W_\ve{z} \ve{z}_k^{(p)} \Vert_{\infty}$
is convex and so is $\mathcal{\tilde{J}}_\ve{z} = \max_{k\in \mathcal{K}} \max_{p\in \mathcal{P}} \Vert W_\ve{z} \ve{z}_k^{(p)} \Vert_{\infty} $. Similarly, $\Vert \ve{u}_k \Vert_1$ is convex, and since the sum of convex functions is convex, so is $\mathcal{\tilde{J}}_\ve{u} =  \epsilon \sum_{k\in \mathcal{K}} \Vert \ve{u}_k \Vert_1 $ and the overall objective $\mathcal{\tilde{J}}_\ve{z} + \mathcal{\tilde{J}}_\ve{u}$. Since the absolute value is convex, $\left | \ve{u}_k \right |\leq \overline{\ve{u}}$ and $\left |\ve{u}_{k+1} - \ve{u}_k\right |\leq h \overline{\overline{\ve{u}}}$ are convex. The other constraints are linear and thus convex. While any general purpose convex solver could be used to solve problem \eqref{eq:mainPb}, elementary transformations can be used to recast it into an equivalent linear optimisation problem for which dedicated and efficient solvers exist. This is demonstrated in the next section.

\subsection{Equivalent linear problem}
\label{ssec:equivalent_lp}
Problem \eqref{eq:mainPb} is not only convex, but also composed solely of either affine or piecewise affine functions and can therefore be recast as an equivalent linear optimisation problem of the form,
\begin{equation}
   \begin{array}{rc}
     \displaystyle \min_{\boldsymbol{\xi}} & \ve{c}^T \boldsymbol{\xi}\\
     s.t.&
     \begin{array}[t]{rcl}
         A_i \boldsymbol{\xi}&\leq&\ve{b}_i\\
         A_e \boldsymbol{\xi}&=&\ve{b}_e \\
         \boldsymbol{\xi}&\in&[\underline{\boldsymbol{\xi}},\overline{\boldsymbol{\xi}}]
     \end{array}
   \end{array}
   \label{eq:lin_pb}
\end{equation}
To transform Problem \eqref{eq:mainPb} into \eqref{eq:lin_pb}, simple reformulation techniques as the ones described in \cite[chap.4]{boyd:convex:2004} are used. In particular,
\begin{itemize}
    \item the absolute values in the constraints are directly replaced with linear constraints as follows,
     \begin{equation}
         \left | \ve{u}_k \right |\leq \overline{\ve{u}} \Longleftrightarrow -\overline{\ve{u}} \leq \ve{u}_k \leq \overline{\ve{u}},
     \end{equation}
     and 
        \begin{equation}
            \left |\ve{u}_{k+1} - \ve{u}_k\right |\leq h \overline{\overline{\ve{u}}}\Longleftrightarrow 
            -h\overline{\overline{\ve{u}}} \leq \ve{u}_{k+1} - \ve{u}_k\leq h\overline{\overline{\ve{u}}}. 
            \label{eq:urate}
        \end{equation}
    \item $\mathcal{\tilde{J}}_\ve{z}$ is replaced by a slack variable $z_s\in \rs_+$ such that $\forall k \in \mathcal{K}$, $\forall p \in \mathcal{P}$, $\forall i= 1,\ldots,n_z$,
    \begin{equation}
        -z_s \leq \ve{e}_i^T W_\ve{z} \ve{z}_k^{(p)}\leq z_s,
        \label{eq:zs}
    \end{equation}
    where $\ve{e}_i \in \mathbb{R}^{n_z}$ is the $i$-th canonical vector. This slack variable lies in the epigraph of $\mathcal{\tilde{J}}_\ve{z}$ and matches it at the optimum.
    \item Similarly, $\mathcal{\tilde{J}}_\ve{u}$ is replaced by $\ve{c}_u^T \ve{u}_s$ where $ \ve{c}_u=\epsilon \mat{ccc}{1&\ldots&1}^T \in \rs[n_u N]$ and where the slack variable $\ve{u}_s \in \rs[n_u N]_+$ must satisfy,
    \begin{equation}
        -\ve{u}_s \leq \mat{c}{\ve{u}_0\\\vdots\\\ve{u}_{N-1}} \leq \ve{u}_s.
        \label{eq:us}
    \end{equation}
    At the optimum, the entries of the slack variable $\ve{u}_s$ will be equal to the absolute values of the entries of the command inputs $\ve{u}_k$, $k\in \mathcal{K}$ so that $\ve{c}_u^T \ve{u}_s$ is equal to $\mathcal{\tilde{J}}_\ve{u}$.
\end{itemize}
The augmented optimisation variable $\boldsymbol\xi \in \mathbb{R}^{n_\xi}$ in \eqref{eq:lin_pb} is then obtained by stacking all the variables\footnote{excepted $\ve{x}_{N}$ which is actually not exploited. The initial state $\ve{x}_0$ could also be discarded as optimisation variable since it is set as $\ve{x}(0)$.} of Problem \eqref{eq:mainPb} and the slack variables $z_s$ and $\ve{u}_s$, e.g.
\begin{equation}
    \small
    \boldsymbol \xi = 
    \left [ \ve{vec}([\underbrace{\ve{u}_0\ldots\,\ve{u}_{N-1}}_{\#n_u N}])^T\, \ve{vec}([\underbrace{\ve{x}_0^{(1)}\ldots\,\ve{x}_{N-1}^{(n_p)}}_{\#n_x n_p N}])^T\,
     \ve{vec}([\underbrace{\ve{z}_0^{(1)}\ldots\,\ve{z}_{N-1}^{(n_p)}}_{\#n_z n_p N}])^T\,
    \underbrace{z_s}_{\#1}\,
    \underbrace{\ve{u}_s^T}_{\#n_u N} \right ]^T.
    \label{eq:xi}
\end{equation}
The matrices of linear inequality constraints $A_i\in \mathbb{R}^{(2 n_u N + 2 n_z n_p N +2 n_u N)\times n_\xi}$ and equality constraints $A_e\in\mathbb{R}^{(n_x  n_p N+ n_z  n_p N)\times n_\xi}$ must then be filled accordingly to the location of each variable in $\boldsymbol \xi$. In particular, let us split the slack vector $\ve{u}_s$ so that \eqref{eq:us} becomes, for $k \in \mathcal{K}$, $   -\ve{u}_{k,s} \leq  \ve{u}_k \leq \ve{u}_{k,s}$, then the pattern of the inequalities is as follows,
\begin{equation}
\resizebox{\linewidth}{!}{$%
\begin{array}{cc} 
    &\begin{array}{*{11}{Z}Z} 
\color{gray}       \ldots & \color{gray}\ve{u}_k &\color{gray}\ve{u}_{k+1} & \color{gray}\ldots& \color{gray}\ldots&\color{gray}\ve{z}_k^{(p)}&\color{gray}\ldots&\color{gray} z_s &\color{gray}\ldots &\color{gray}\ve{u}_{k,s} &  \color{gray}\ldots &\color{gray}\ve{b}_i\\
    \end{array}
    \\
    \left [ A_i \vert \ve{b}_i \right ]=&
    \left[
    \begin{array}{*{11}{Z}|Z} 
    \ddots&\ddots &  & &&&&&&&&\vdots\\
          & -I_{n_u}& I_{n_u} & &&&&&&&&h \overline{\overline{\ve{u}}}\\
        & &\ddots& \ddots & &&&&&&&\vdots\\
            \ddots&\ddots &  & &&&&&&&&\vdots\\
          & I_{n_u}& -I_{n_u} & &&&&&&&&h\overline{\overline{\ve{u}}}\\
        & &\ddots& \ddots & &&&&&&&\vdots\\\hline
        & && &\ddots  &&&\vdots&&&&\vdots\\
        & && & & W_\ve{z}&&-\mathbb{1}_{n_z}&&&&0\\   
        & && &&&\ddots  &\vdots&&&&\vdots\\
        & && &\ddots  &&&\vdots&&&&\vdots\\
        & && & & -W_\ve{z}&&\mathbb{1}_{n_z}&&&&0\\   
        & && &&&\ddots  &\vdots&&&&\vdots\\\hline
    \ddots& &  & &&&&&\ddots&&&\vdots\\
        & I_{n_u}&& & &&&&&-I_{n_u}&&0\\   
                &&\ddots& & &&&&&&\ddots&\vdots\\   
        \ddots&&& & &&&&\ddots&&&\vdots\\   
        & -I_{n_u}&& & &&&&&I_{n_u}&&0\\   
        & &\ddots&  & &&&&&&\ddots&\vdots\\
    \end{array}
    \right]
\end{array}%
$}
\label{eq:ai}
\end{equation}

where $I_n\in\mathbb{R}^{n\times n}$ is the identity matrix and $\mathbb{1}_n \in \mathbb{R}^n$ is a column vector filled with ones. The variables associated with the different columns are indicated at the top in grey. The different horizontal partitions in \eqref{eq:ai} are associated with the constraints \eqref{eq:urate}, \eqref{eq:zs} and \eqref{eq:us}, respectively. Similarly, the pattern for the equalities is as follows,
\begin{equation}
\resizebox{\linewidth}{!}{$%
\begin{array}{cc} 
    &\begin{array}{*{10}{Z}W} 
        \color{gray} \ldots &\color{gray} \ve{u}_k  &\color{gray} \ldots& \color{gray}\ldots&\color{gray}\ve{x}_k^{(p)}&\color{gray} \ve{x}_{k+1}^{(p)}&\color{gray}\ldots&\color{gray}\ldots&\color{gray} \ve{z}_k^{(p)}&\color{gray} \ldots & \color{gray}  \ve{b}_e\\
    \end{array}
    \\
    \left [ A_e \vert \ve{b}_e \right ]=&
    \left[
    \begin{array}{*{10}{Z}|W} 
    \ddots&  & &\ddots&\ddots&&&&&&\vdots\\
    &-\mathcal{B}_2 & & &-\mathcal{A}&I_{n_x}&&&&&\mathcal{B}_1 \ve{w}_k^{(p)}\\
    &  &\ddots &&&\ddots&\ddots&&&&\vdots\\\hline
    \ddots& &&\ddots&&&&\ddots&&&\vdots\\
    & -\mathcal{D}_{12} &&&-\mathcal{C}_1&&&&I_{n_z}&&\mathcal{D}_{11} \ve{w}_k^{(p)}\\
    & &\ddots&&&\ddots&&&&\ddots&\vdots\\
    \end{array}
    \right]
\end{array}%
$}
\end{equation}
where the first partition is associated with the dynamical equation and the second one with the performance output.

\subsection{Decreasing the number of optimisation variables}
\label{ssec:nvars}

The number of optimisation variables $n_\xi $ is equal to $ ( n_u + n_x n_p + n_z n_p ) N  + 1$ without the slack variables associated with $\tilde{\mathcal{J}}_\ve{u}$ which adds $n_u N$ variables if considered. This number can grow quite fast depending on the dimensions of the dynamical model. While very large linear problems can be solved efficiently with dedicated solvers, the dimension can still become detrimental should it be only in terms of resolution time. Before relying on the simplification processes embedded in linear solvers, several actions may be considered to reduce the number of variables and constraints. 

First, from equation \eqref{eq:dsys}, note that $\forall k \in \mathcal{K}$, $\forall p \in \mathcal{P}$, the variables $\ve{z}_k^{(p)}$ are linear combinations of $\ve{x}_k^{(p)}$, $\ve{w}_k^{(p)}$ and $\ve{u}_k$ and can therefore be discarded as variables and replaced directly in the inequality constraints by these linear combinations. This enables to save $n_z n_p N$ variables and associated equality constraints.

\paragraph{Adjusting the shooting scheme. }Going further in that direction, note that the intermediary states $\ve{x}_k^{(p)}$ can also be discarded. The performance output at time $k$ can directly be  written as a linear function of the past control inputs and disturbances. More specifically, $\forall k \in \mathcal{K}$, $\forall p \in \mathcal{P}$,
\begin{equation}
    \ve{z}_k^{(p)} = \mathcal{C}_1 \left ( \mathcal{A}^k \ve{x}_0 + \sum_{j=0}^{k-1} \mathcal{A}^{k-j-1} (\mathcal{B}_1 \ve{w}_j^{(p)}+ \mathcal{B}_2 \ve{u}_j) \right ) + \mathcal{D}_{11} \ve{w}_k^{(p)} + \mathcal{D}_{12} \ve{u}_k.
    \label{eq:condensed_out}
\end{equation}
It is therefore possible to discard completely the intermediate variables $\ve{x}_k^{(p)}$ and $\ve{z}_k^{(p)}$ by replacing each occurrence of the latter by \eqref{eq:condensed_out}. While this approach enables to discard $n_x n_p N$ additional optimisation variables (as well as the remaining equality constraints), care must be taken as computing large power of matrices can lead to some severe numerical issues and bad conditioning of the resulting matrix of inequality constraints. From a theoretical perspective, this approach can be seen as a single shooting method \cite{pesh:practical:1994} and a trade-off can be obtained by conserving only some intermediate state variables which would result in a multiple shooting scheme.

\paragraph{Model order reduction. }Should (some) intermediary states be kept as variables, then model order reduction (\cite{antoulas:approximation:2005}) is a complementary way to decrease the number of associated variables. Considering a LTI model such as \eqref{eq:dsys}, model order reduction is generally aimed at finding a reduced model $\hat{H}$ of dimension $r \ll n$ that well reproduces the input-to-output behaviour of $H$ in some sense. The $\mathcal{H}_2$-norm is commonly exploited to quantify the induced error. Indeed, among its interesting properties it enables to bound globally the worst output error. In particular, let the augmented input of the model \eqref{eq:dsys} obtained by concatenating the control input $\ve{u}$ with the disturbance input $\ve{w}$ be denoted by $\tilde{\ve{u}}$, then for all input signal $\tilde{\ve{u}} \in \mathcal{L}_2$ of bounded energy, the worst output error induced by reduction is bounded as follows,
\begin{equation}
    \Vert \ve{z} - \ve{\hat{z}} \Vert_\infty \leq \underbrace{\Vert H - \hat{H} \Vert_2 \Vert \tilde{\ve{u}} \Vert_2}_{=\gamma(\tilde{\ve{u}})}.
    \label{eq:bound}
\end{equation}
Provided that the approximation error is small enough, the bound \eqref{eq:bound} suggests that the output error should also be small. Obviously, the approximation error depends on the approximation order $r$ and on the characteristics of the model. As detailed by \cite{antoulas:approximation:2005}, fast decaying Hankel singular values indicates that a model can be approximated accurately with a low-order model.

Although it may be conservative in practice (see section \ref{sec:appli}), the bound \eqref{eq:bound} can be integrated in problem \eqref{eq:init_pb} as an additive uncertainty to ensure the robust performance of the sought control signal. The idea consists in replacing the large-scale model output $\ve{z}$ by the reduced-order one $\hat{\ve{z}} $ completed by an uncertainty $\boldsymbol \rho$ with $ \Vert \boldsymbol \rho \Vert_\infty \leq \gamma(\tilde{\ve{u}}) $ so that the optimal control problem \eqref{eq:init_pb} becomes,
\begin{equation}
    \sup_{\Vert \boldsymbol{\rho} \Vert_\infty \leq \gamma(\tilde{\ve{u}})} \max_{\hat{\ve{z}} \in \hat{\mathcal{Z}}(\ve{u}, \mathcal{W})}  \Vert W_z (\hat{\ve{z}} + \boldsymbol{\rho}) \Vert_\infty,
    \label{eq:robustpb}
\end{equation}
where $\hat{\mathcal{Z}}(\ve{u}, \mathcal{W})$ is the set of performance output obtained through the reduced model $\hat{H}$. In the discrete time-domain, this translates into the following problem,
\begin{equation}
  \begin{array}{rc}
     \displaystyle \min_{\left \{ \ve{u}_k\right \}_{k\in \mathcal{K}}} & \displaystyle   \sup_{\Vert \boldsymbol{\rho} \Vert_\infty \leq \gamma(\tilde{\ve{u}})} \max_{k\in \mathcal{K}} \max_{p\in \mathcal{P}} \Vert W_\ve{z} (\hat{\ve{z}}_k^{(p)} + \boldsymbol{\rho}) \Vert_{\infty} + \epsilon \sum_{k\in \mathcal{K}} \Vert \ve{u}_k \Vert_1 
     \\
     s.t.\, \forall k \in \mathcal{K},\,\forall p\in \mathcal{P},
     &
     
     \begin{array}[t]{rcl}
        \hat{\ve{x}}_{k+1}^{(p)}&=&\hat{\mathcal{A}} \hat{\ve{x}}_k^{(p)} + \hat{\mathcal{B}}_{1} \ve{w}_k^{(p)}+ \hat{\mathcal{B}}_2\ve{u}_k\\
        \hat{\ve{z}}_{k}^{(p)}&=&\hat{\mathcal{C}}_1 \hat{\ve{x}}_k^{(p)} + \hat{\mathcal{D}}_{11} \ve{w}_k^{(p)}+ \hat{\mathcal{D}}_{12}\ve{u}_k\\
        \left | \ve{u}_k \right |&\leq& \overline{\ve{u}}\\
        \left |\ve{u}_{k+1} - \ve{u}_k\right |&\leq&h\overline{\overline{\ve{u}}}\\
        \hat{\ve{z}}_k^{(p)} + \boldsymbol{\rho}&\leq& \overline{\ve{z}}\\
        \hat{\ve{z}}_k^{(p)} + \boldsymbol{\rho}&\geq& \underline{\ve{z}}\\
        \hat{\ve{x}}_0^{(p)} &=&\hat{\ve{x}}(0)
     \end{array}
  \end{array}
  \label{eq:robustPb}
\end{equation}
This problem remains convex but is no longer equivalent to a linear program due to the constraint $\Vert \boldsymbol{\rho} \Vert_\infty \leq \gamma(\tilde{\ve{u}})$ which includes a quadratic element in the right hand side. Suppose an upper bound $\overline{\gamma} \geq \gamma(\tilde{\ve{u}})$ is available then the problem can be transformed to an equivalent linear problem again at the cost of an increased conservatism.

Indeed, as detailed by \cite{vandenberghe:2002:robustlp}, inverting the sup with the max and introducing the slack variable $z_s$ leads to the constraints $\forall k \in \mathcal{K}$, $\forall p\in \mathcal{P}$, $i =1,\ldots,n_z$,
\begin{equation}
    \sup_{\Vert \boldsymbol{\rho} \Vert_\infty \leq\overline{\gamma}}\vert \ve{e}_i^T W_\ve{z} (\hat{\ve{z}}_k^{(p)} + \boldsymbol{\rho}  ) \vert\leq z_s,
\end{equation}
which is true if and only if 
\begin{equation}
    - z_s + \Vert \ve{e}_i^T W_k \Vert_1 \overline{\gamma} \leq \ve{e}_i^T W_z \hat{\ve{z}}_k^{(p)} \leq z_s - \Vert \ve{e}_i^T W_k \Vert_1 \overline{\gamma}.
\end{equation}
The same process, which may be considered as a constraint tightening approach, is also applied to the constraints on the output performances that should be replaced with 
\begin{equation}
    \underline{\ve{z}} + \overline{\gamma} \leq \hat{\ve{z}}_k^{(p)} \leq \overline{\ve{z}} - \overline{\gamma}.
    \label{eq:robustperf}
\end{equation}
Note that unlike the full-order case, the overall optimisation problem may be infeasible if the reduced-order model is too far from the full-order one. Indeed, $\ve{u} = 0$ does not necessarily enable to satisfy the robust performance constraint \eqref{eq:robustperf}.

\section{Numerical illustration}
\label{sec:appli}

This section is aimed at illustrating the optimal control approach on a realistic\footnote{For practical loads alleviation control purposes, linear models are generally considered as they are representative enough. See for instance the industrial application \cite{poussot:2020:gla}.} aeroelastic aircraft model. The latter is described in section \ref{ssec:model} together with its reduced-order counterpart and the limits of the robust approach described above. Then, in section \ref{ssec:perfs}, the performances of the aircraft with respect to gust-load alleviation are analysed.

\begin{figure}
    \centering
    \includegraphics[width=\linewidth,keepaspectratio]{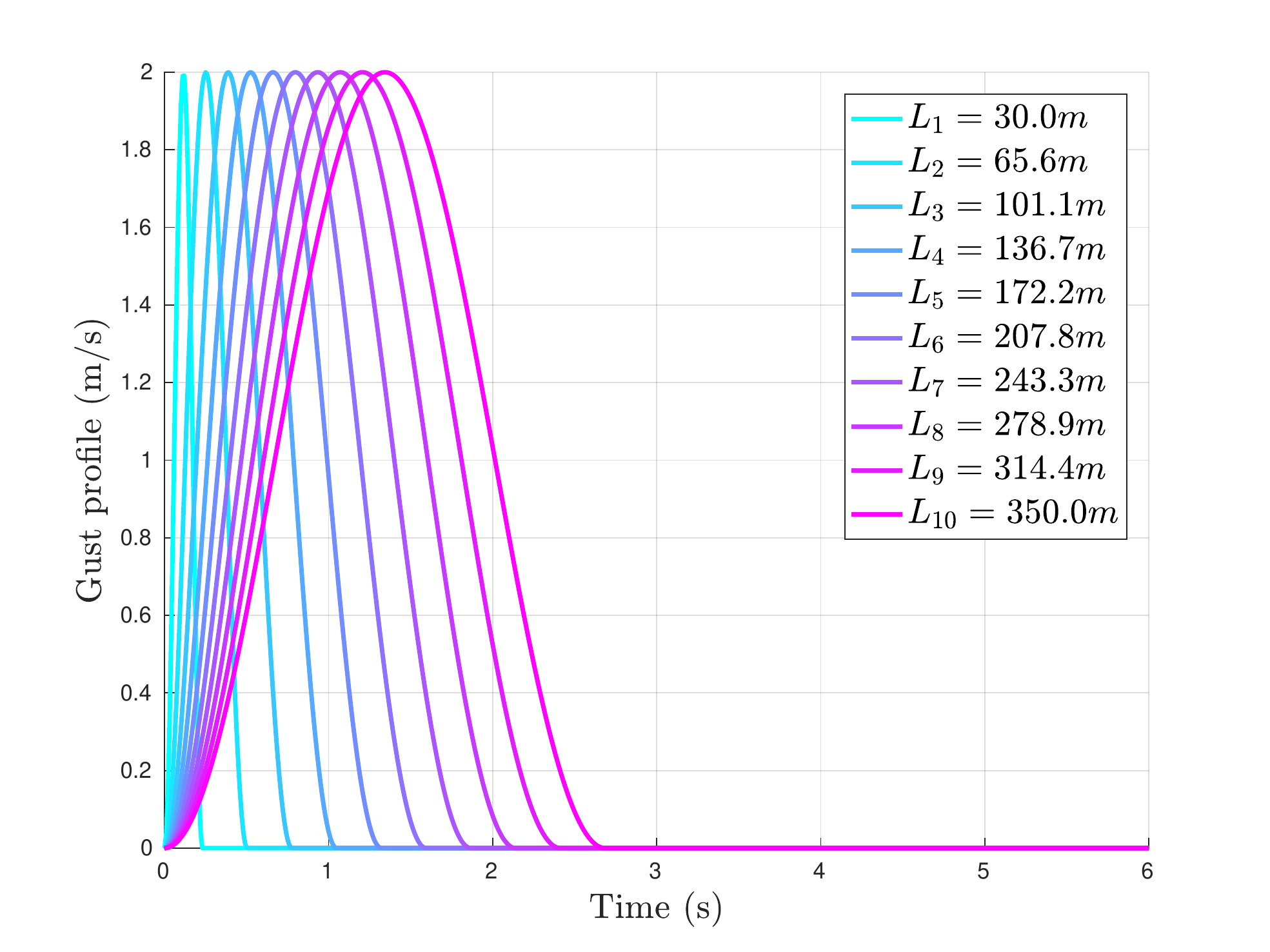}
    \caption{$1-cosine$ gust profiles with wavelength spread from $30m$ to $350m$.}
    \label{fig:gusts}
\end{figure}

The $1-cosine$ gust disturbances used throughout this section have the following form,
\begin{equation}
    w(t) = \left \lbrace 
    \begin{array}{ll}
    \frac{W}{2} (1 - cos( \frac{\pi V}{L} t)) & \text{ if } t \leq 2\frac{L}{V}\\
    0&\text{ otherwise.}
    \end{array}
    \right .
\end{equation}
where $W$ and $L$ are the velocity and wavelength of the gust and $V$ is the aircraft true airspeed. Several values of these parameters are provided by the certification authorities. In the sequel, $n_p = 10$ gust profiles are considered with wavelength $L$ spread within $[30,150]m$ and with velocity $W=2m/s$ thus leading to the profiles in figure \ref{fig:gusts}. Smaller wavelength leads to higher frequency profiles with less energy.

\subsection{Aeroelastic aircraft modelling and reduction}
\label{ssec:model}

The considered model represents the linearised behaviour of a generic flexible long range aircraft flying at $9100m$ and Mach number $0.86$ (thus $V = 260.58m/s$). The state-space realisation is obtained through interpolation of frequency-domain data generated from a subsonic potential flow aerodynamic code following the approach described by \cite{aerospace6010009}. 

The initial model has $16$ inputs gathering the gust input and the control inputs: inner and outer ailerons on each wing, the elevator and the derivatives of these quantities. The gust input represents a vertical disturbance which induces a symmetric response of the aircraft. The latter is handled in practice by a symmetric control action as illustrated by the industrial feedback design application by \cite{poussot:2020:gla}. Thus, left and right ailerons are coupled. In addition, second order low-pass models are added to each control input to represent the dynamic of the actuators. This enables to filter out the derivatives actions as they become negligible in the input-output energy transfer. 

The initial model contains $2527$ outputs containing flight dynamic quantities, various bending moments and torsion along the left and right wings. A subset of $5$ bending moments are selected on each wing and added to exploit the symmetries of the problem. 

All in all, the resulting model $H$ is a $270$-th order model with $n_u = 4$ inputs: the gust, the inner ailerons, outer ailerons and the elevator and $n_z = 5$ performance outputs representing bending moments from the wing root up to $23m$ on the wing. Note that the wing is $30m$ long. Bending moments further to the tip are extremely small and act through the constraint \eqref{eq:perf_cst} as the main limiting factor to a point where they can prevent from any action. This can make the results confusing which is why we limit the analysis up to $23m$ where the largest loads are encountered.

\begin{figure}
    \centering
    \includegraphics[width=\linewidth]{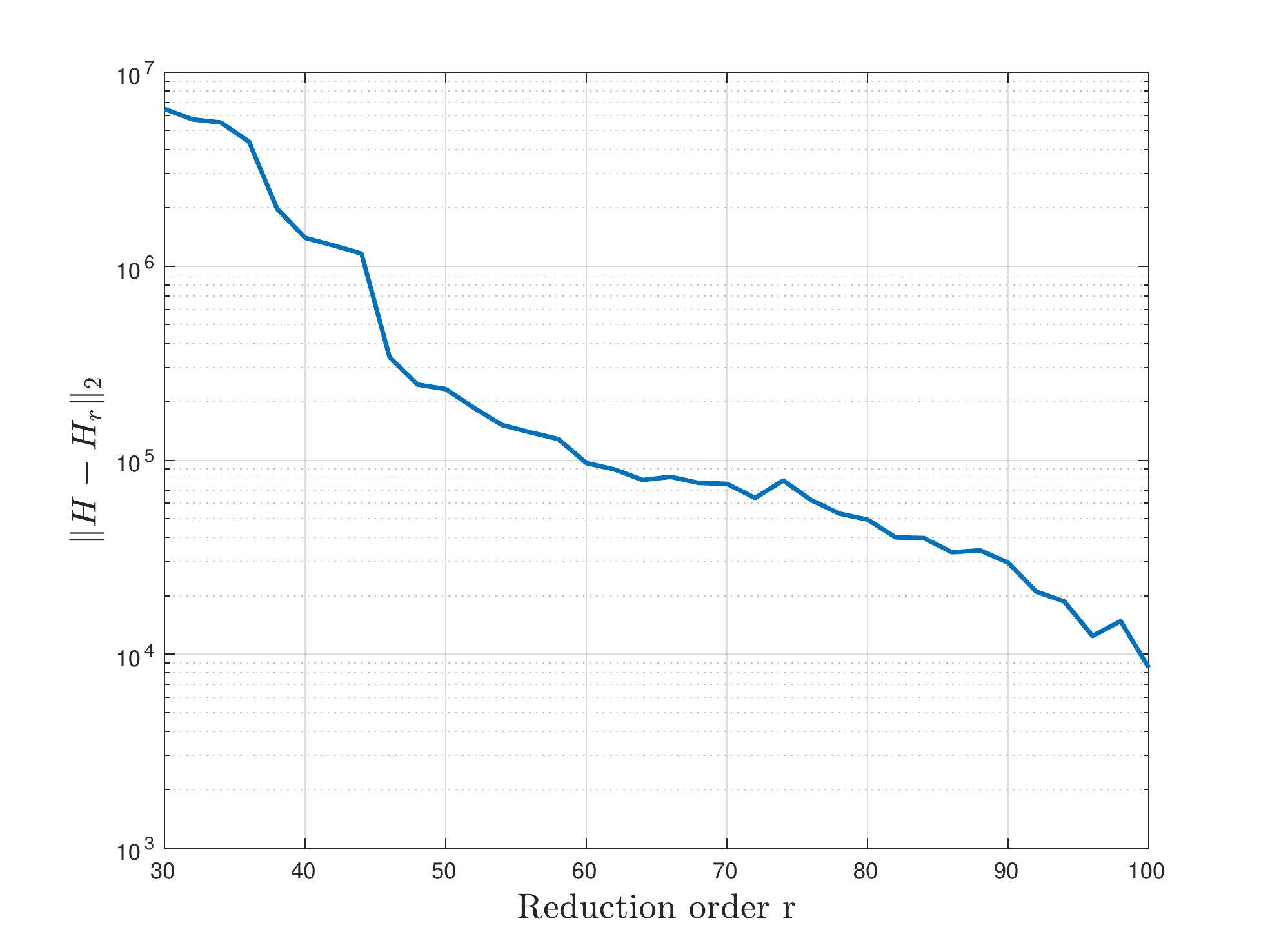}
    \caption{$\mathcal{H}_2$-norm of the approximation error against the approximation order $r$ ranging from $30$ to $100$.}
    \label{fig:red_error}
\end{figure}

\begin{figure}
    \centering
    \includegraphics[width=0.8\linewidth]{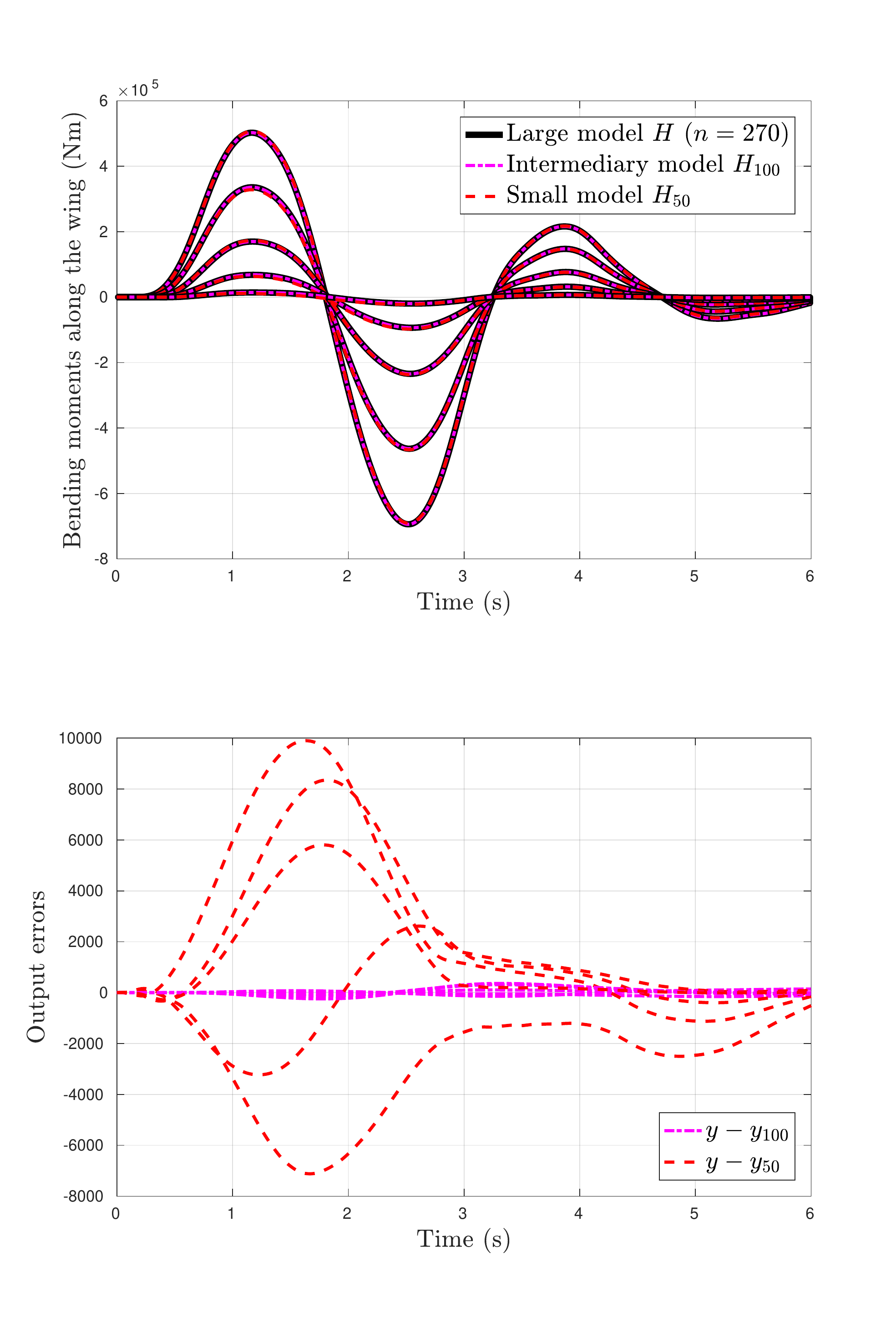}
    \caption{Output signals (left) and associated errors (right) when the large-scale aeroelastic model and its reduction at orders $100$ and $50$ are excited by a $1-cosine$ gust profile.}
    \label{fig:red_output}
\end{figure}

\paragraph{Reduced-order modelling. }The large-scale aeroelastic model $H$ is reduced with a local $\mathcal{H}_2$-optimal model reduction method (\cite{VanDooren:2008}) for varying approximation orders. The associated approximation errors are reported in figure \ref{fig:red_error}. While the absolute error may appear very large, note that $\Vert H \Vert_2 = 2.77\cdot 10^8$ meaning that for $r=30$ (first point of the curve), the relative error is only of $2.33 \%$. In figure \ref{fig:red_output}, the output bending moments obtained when the large-scale model is excited with the largest gust profile without any input command is compared with the ones obtained with two reduced-order models $H_{100}$ and $H_{50}$ which subscripts indicate the order.

This open-loop comparison enables to put the error bound \eqref{eq:bound} in perspective. Indeed, for the considered gust profile\footnote{The situation is slightly better with the smallest gust profile for which $\Vert w \Vert_2 = 6$.} $\Vert \tilde{\ve{u}} \Vert_2 = \Vert \ve{w} \Vert_2 \approx 20$ which means that 
\begin{equation}
\Vert \ve{y} - \ve{y}_{100} \Vert_\infty \leq  \Vert H - H_{100} \Vert_2 \Vert \tilde{\ve{u}} \Vert_2 \approx 1.7 \cdot  10^5,
\label{eq:b1}
\end{equation}
and 
\begin{equation}
\Vert \ve{y} - \ve{y}_{50}\Vert_\infty \leq  \Vert H - H_{50} \Vert_2 \Vert \tilde{\ve{u}} \Vert_2 \approx 4.7 \cdot 10^6.
\label{eq:b2}
\end{equation}
With respect to figure \ref{fig:red_output} (bottom), we can see that $\Vert \ve{y} - \ve{y}_{100} \Vert_\infty < 10^3$ and $\Vert \ve{y} - \ve{y}_{50} \Vert_\infty \approx  10^4$. This highlights the conservatism of the bounds \eqref{eq:b1} and \eqref{eq:b2}. In addition, by looking at the magnitude of the bending moments for this gust input in figure \ref{fig:red_output} (top), we can see that the bound \eqref{eq:b1} cannot be exploited for the smallest bending moment output which magnitude is actually lower. The situation is even worse for the $50$-th order model which error bound \eqref{eq:b2} is larger that the largest output bending moment.

These elements highlight the limits of the robust approach mentioned in section \ref{ssec:nvars}. Indeed, for the method to be relevant, a reduced-order model of order larger than $100$ must be considered while an order $50$ already leads to moderate output errors. This counterbalances the potential of the approach to decrease the number of optimisation variables $n_\xi$ as illustrated by table \ref{tab:nvars}. 

\begin{table}[]
    \centering
    \begin{tabular}{c|ccc}
        & $H$   &$H_{100}$  &$H_{50}$   \\ \hline
 $n_\xi$&       $1\,621\,801$     & $      601\,801$          &   $       301\,801$         \\ 
 \end{tabular}
    \caption{Number of variables associated with the optimal control problem for a time horizon of $T = 6s$, a sampling-time $h=0.01s$, $n_p = 10$ gust profiles, no complementary criterion $\mathcal{J}_\ve{u}$ and no variable dedicated to $\ve{z}_k^{(p)}$. Note that within $n_\xi$, the number of variables associated with the control input is $n_u N = 1\,800$.}
    \label{tab:nvars}
\end{table}

The conservatism largely depends on the considered model and its potential for reduction. Other cases may be more suitable to integrate the robust bound \eqref{eq:bound}. Besides, decreasing the number of inputs or outputs may also help decrease even further the approximation error. In the sequel, the reduced-order model $H_{50}$ is used within the optimisation problem without the robust bound and the large-scale model is used afterwards to validate the results in simulation.

Note that the single shooting approach mentioned in section \ref{ssec:nvars} is not considered as it leads to an excessive sensitivity when solving the associated linear optimisation problem. More specifically, no improvement is achieved. 

\subsection{Performances evaluation}
\label{ssec:perfs}

Throughout this section and unless specified otherwise, the following numerical parameters are considered: control horizon is $T=6s$, sampling period is $h = 0.01s$. In addition, the actuators (inner and outer ailerons and elevators) are symmetrically saturated at $15deg$. They are also saturated in rate at $20deg/s$ for the ailerons and $5deg/s$ for the elevator. 

This general setting is adjusted to conduct four experiments in the sequel:
\begin{inlinelist}
\item comparison of the efficiency of each actuator,
\item evaluation of the impact of saturations,
\item evaluation of the impact of some delays in the control action,
\item evaluation of the impact of uncertainty of the disturbance.
\end{inlinelist}

\begin{figure}
    \centering
    \includegraphics[width=\linewidth,keepaspectratio]{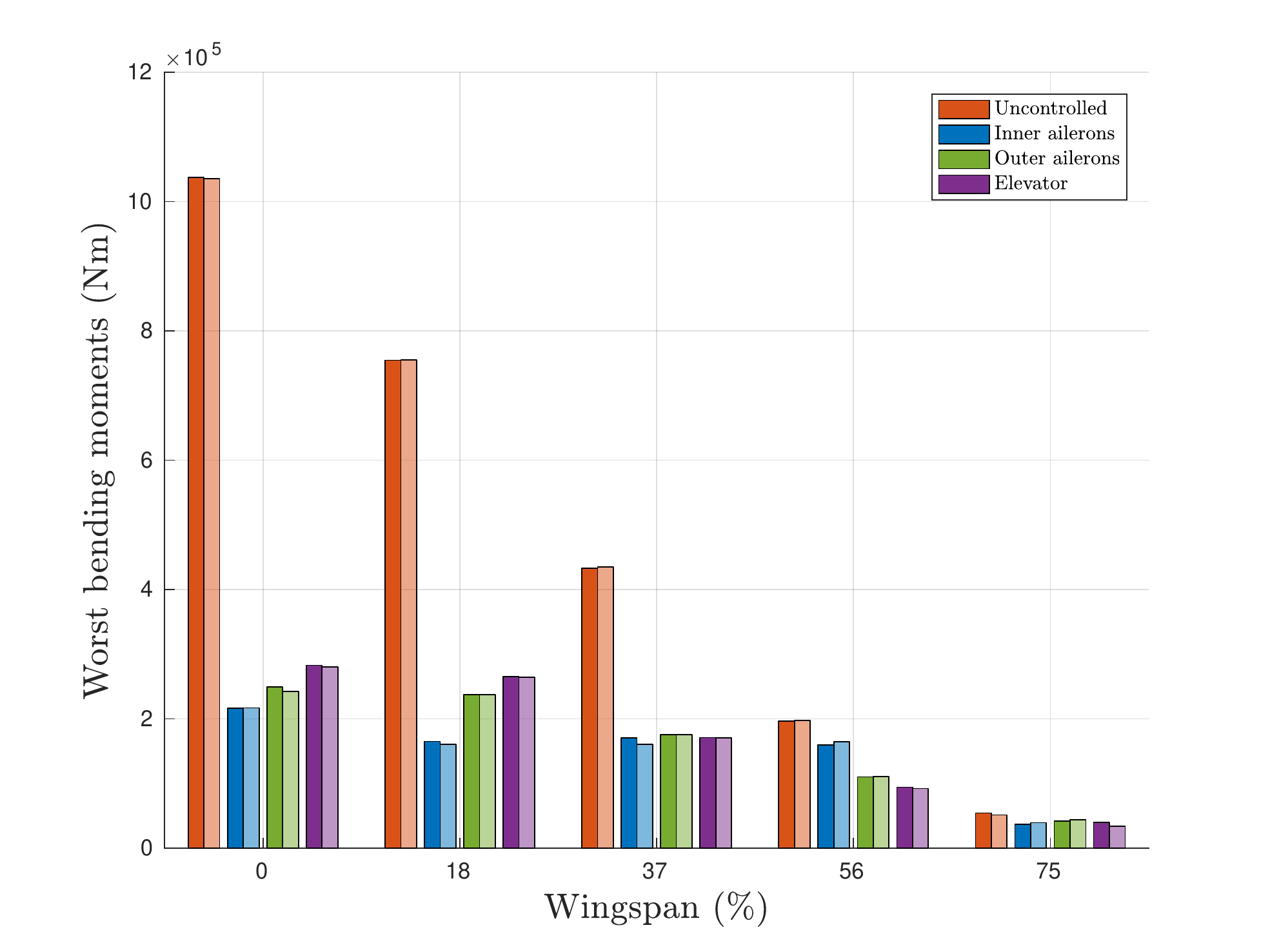}
    \caption{Worst bending moments along the wing without control and using either one of the available actuator: inner ailerons, outer ailerons or elevator. The optimal control problem is solved with the reduced-order model $H_{50}$. Darker values of a color are associated with the bending moments obtained after simulation on the large-scale model $H$ while light values correspond to the expected bending moments from the reduced-order model $H_{50}$.}
    \label{fig:comp_act}
\end{figure}

\paragraph{Comparison of actuators performances. }In order to compare the efficiency of each actuator for gust load alleviation, the optimal control problem is adjusted to consider only one single actuator at a time. The problem is solved for each actuator and each gust profile. The worst bending moments are measured and are reported in figure \ref{fig:comp_act}.

The accuracy of the reduced-order model is confirmed here as there is very little difference between the expected bending moments returned by the (reduced) control problem (lighter colours) and the ones obtained on the large-scale model (darker colours). This holds both in the uncontrolled and controlled cases. Figure \ref{fig:comp_act} also highlights clearly the difference of magnitude between the bending moments observed along the wing. The largest ones are observed near the fuselage and are generally the sizing ones. 

Again, we emphasise the fact that the particularly impressive decrease of the loads must be taken with care as the exact knowledge of the disturbances is considered and no specific dynamic is tied to the control signal. This could hardly be achieved by a feedback control law in practice. Yet, while sharing the same framework, some actuators perform better than others. In particular, with the considered configuration, it appears that inner ailerons are the most efficient followed by the outer ailerons and the elevator.
This could be expected as ailerons are generally the dedicated actuators for gust loads alleviation due to their location on the wing. The elevator has quite a different impact as it induces a pitch motion to counteract the effect of the gust.

This highlights the interest of the overall approach in the perspective of novel aircraft design. Indeed, it enables to quickly obtain a quantitative comparison of different actuators. For instance one could compare different locations, geometries, technologies, etc.

\begin{figure}
    \centering
    \includegraphics[width=\linewidth,keepaspectratio]{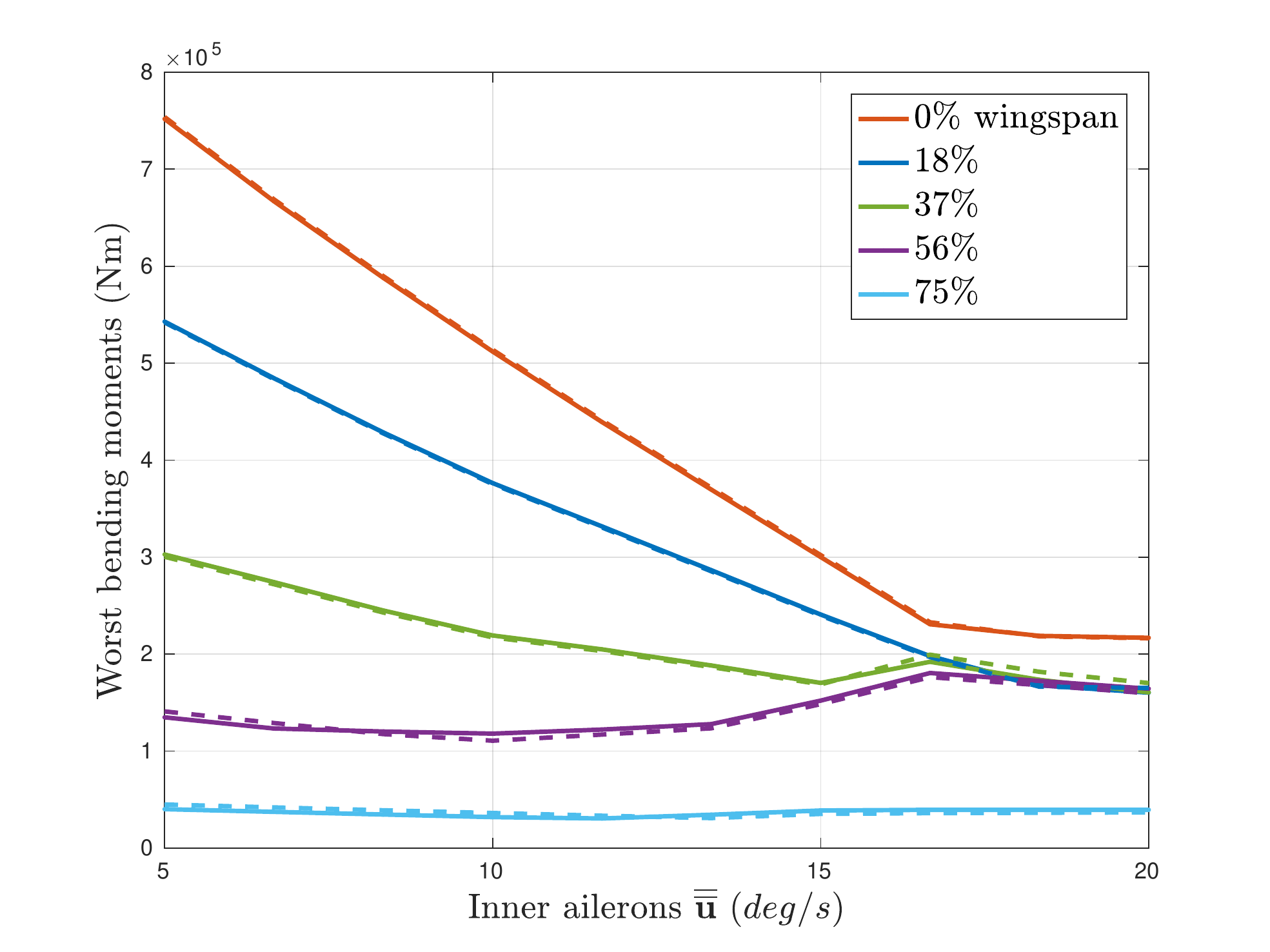}
    \caption{Worst bending moments along the wing against the rate saturation of the inner ailerons when only the latter are used. Plain lines represents the results obtained with the reduced-order model and dashed lines the ones obtained with the full-order model.}
    \label{fig:comp_rate}
\end{figure}

\paragraph{Impact of rate saturation. }The proposed framework also enables to evaluate the impact of some characteristics of the actuators such as their rate limitation. For instance, considering only the inner ailerons, the upper bound $\overline{\overline{\ve{u}}}$ is increased from $5deg/s$ to $20deg/s$ and the resulting worst loads are reported in figure \ref{fig:comp_rate}.

Again, one can note the closeness between the results obtained on the reduced-order model (plain lines) and full-order one (dashed lines). The rate saturation has mainly an impact on the bending moments near the root of the wing which are the sizing cases. As it could be expected, the worst case decreases as the rate saturation increases. It quickly reaches a plateau thus indicating that another constraint then become the limiting factor. On the other hand, the drop in performances is fast below this value thus showing that the rate saturation is indeed a major element for gust load alleviation.

This comparison can in addition be put in parallel to the analysis of the impact of delays on the performances. Indeed, a faster actuator may be able to compensate for larger delays.

\begin{figure}
    \centering
    \includegraphics[width=\linewidth,keepaspectratio]{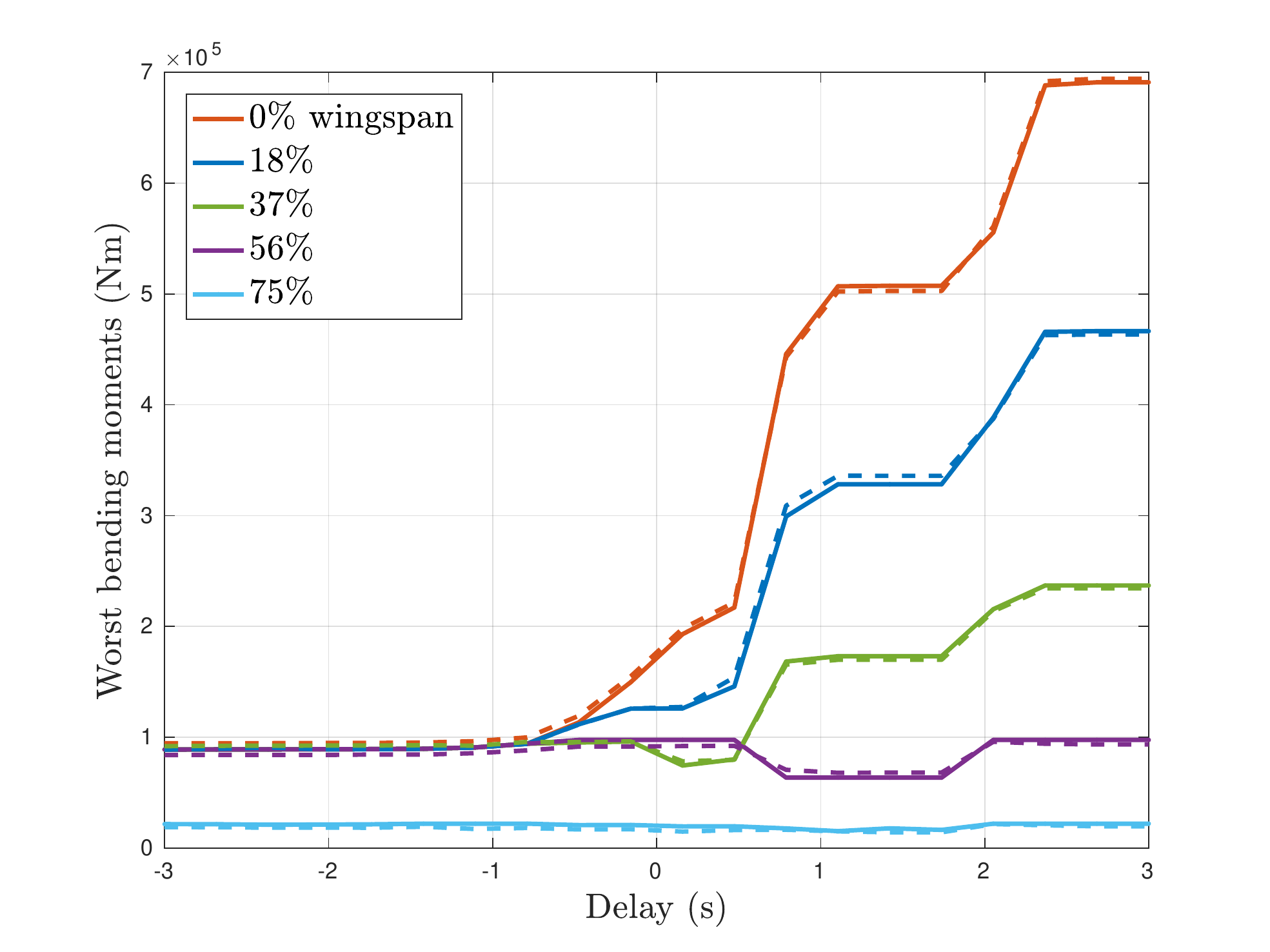}
    \caption{Worst bending moments along the wing against the delay before action when only the inner ailerons are used. Negative delay implies a pre-emptive action. Plain lines represent the results obtained with the reduced-order model and dashed lines the ones obtained with the full-order model.}
    \label{fig:delay}
\end{figure}

\paragraph{Impact of delays. }As detailed in section \ref{sec:optimal_control}, by letting $\overline{\ve{u}}$ vary with time, one can simulate a pre-emptive action or some delay between the moment the gust reaches the aircraft and the beginning of the control action. Let the control action of the inner ailerons be delayed by a value $\tau \in \mathbb{R}$ ranging from $-3s$ to $3s$ (which corresponds to the length of the largest gust profile). A negative value indicates that the control signal may pre-compensate for the effect of the gust. The corresponding worst bending moments along the wing are reported in figure \ref{fig:delay}. 

As expected, the performances deteriorate quite fast as the delay increases. This highlights the importance of a fast control chain for the problem of loads alleviation. Thanks to the inertia of the system, one can notice that some gains can still be achieved with $3s$ of delay that is to say after the gust has completely passed the nose of the aircraft. The evolution by steps corresponds to the oscillations of the bending moments (see e.g. figure \ref{fig:red_output}).

Conversely, figure \ref{fig:delay} shows that a pre-emptive action can help to slightly improve the performances. However, a minimum is reached below $-0.7s$ thus indicating that there is little interest in trying to forecast the disturbance further away. These simulations suggest that the priority should be to minimise the delays in the control chain as it has a major impact.

\begin{figure}
    \centering
    \includegraphics[width=\linewidth,keepaspectratio]{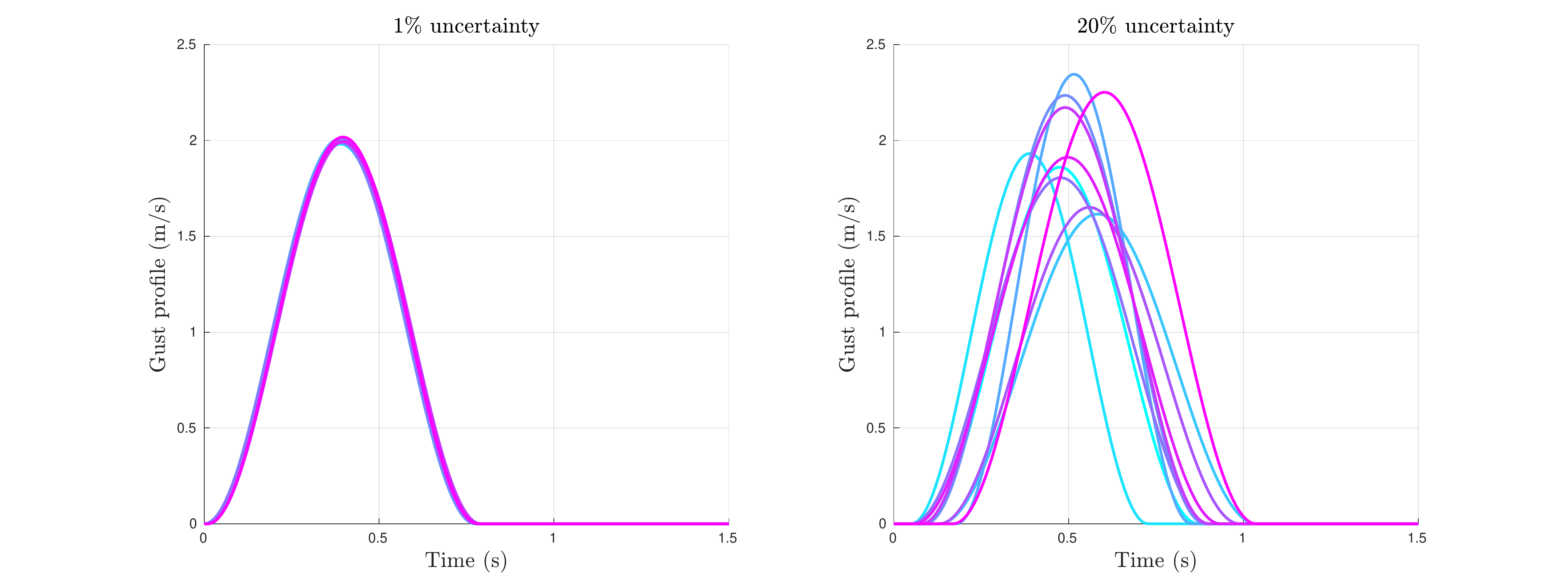}
    \caption{Illustration of the uncertainties added to the gust profile $L_3 = 101.1m$. On the left, the parameters are disturbed at most by $1\%$ around their nominal values and by $20\%$ on the right.}
    \label{fig:ugust}
\end{figure}

\begin{figure}
    \centering
    \includegraphics[width=\linewidth,keepaspectratio]{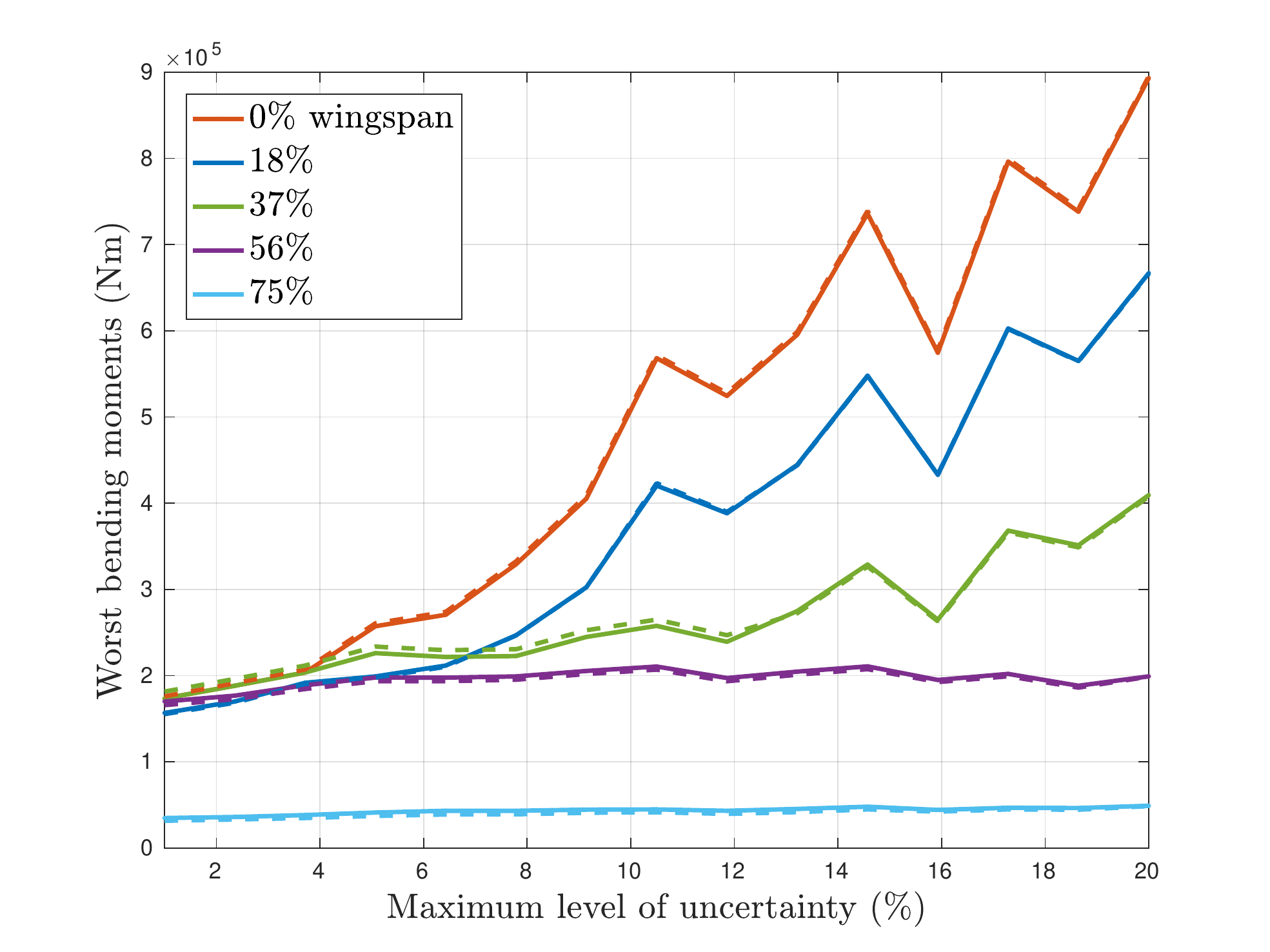}
    \caption{Worst bending moments obtained along the wingspan against the maximum level of uncertainty injected into the disturbance gust profile $L_3 = 101.1m$. For each level of uncertainty, the problem is solved $5$ times by drawing $10$ random gust profiles modelling the uncertainty. Only inner ailerons are considered. Plain lines represent the results obtained with the reduced-order model and dashed lines the ones obtained with the full-order model.}
    \label{fig:uncertainty}
\end{figure}

\paragraph{Uncertainty in the disturbance. }All the previous experiments have been conducted assuming perfect knowledge of the disturbance. This results in a very efficient rejection and very small controlled loads. The knowledge of the shape and date of the gust has a major impact. To highlight this point, let us consider the gust profile with wavelength $L_3 = 101.1m$ which leads to the worst loads. Its amplitude, wavelength and starting date are disturbed by an increasing maximum percentage ranging from $1\%$ to $20\%$ of their nominal values. In figure \ref{fig:ugust}, $10$ gust profiles are randomly drawn for the corresponding maximum level of uncertainty. To mitigate the effect of randomness and to compensate for the moderate number of random profiles, the problem is solved $5$ times for each uncertainty level and the worst cases are retained. The results are reported in figure \ref{fig:uncertainty}. Note that the resulting command input is also tested against the nominal gust profile and the resulting bending moments are included in figure \ref{fig:uncertainty} but this does not modify the curves.

Although non-monotonic due to the randomness of the experiment\footnote{Some combinations of parameters for the gust profiles are particularly detrimental for the loads in comparison to others. Each level uncertainty expands the set of possible gusts and the loads should thus be increasing.}, the loss of performances is clear as the uncertainty level increases. Indeed, for $1\%$ uncertainty, performances similar to the ones observed in \ref{fig:comp_act} are observed. But then, for $20\%$, the $3$ largest bending moments reach about $80\%$ of the uncontrolled loads. The loads at the tip of the wing are barely affected as they are already close to their uncontrolled values. Such levels of performances remain high, but much closer to what could realistically be achieved \cite{poussot:2020:gla}. This experiment highlights the importance of an accurate estimation of the disturbance should any feed-forward control scheme be used for gust load alleviation. And in particular, it can help to assess the level of accuracy required for some sensors such as LIDAR in order to be able to decrease the loads.

\section{Conclusion}
\label{sec:ccl}

In this article, the problem of gust loads alleviation is addressed from an early design perspective with an open-loop optimal control approach. More specifically, considering the linearised behaviour of an aeroelastic aircraft, a direct transcription approach is used to translate the main requirements encountered in practice into an equivalent linear optimisation problem. Practical approaches are described to alleviate the dimension issue of the resulting optimisation problem. In particular, the number of variables can be decreased by changing the shooting scheme at the cost of an increased sensitivity and/or by exploiting model order reduction at the cost of some uncertainty. The overall framework is applied on a realistic aeroelastic model. 

The example highlights the relevance of model reduction in the context of (direct) optimal control. Indeed, it enables to decrease significantly the number of associated optimisation variables while preserving a high level of accuracy. However, the considered bound on the approximation error has proven to be too conservative with this model to be integrated in the optimisation problem. Considering a time-limited approximation method by \cite{goyal:2019:time} is an interesting lead to tighten this bound as the underlying framework matches finite-horizon optimal control. On a more general level, it is still not clear to what extent model reduction could be exploited in the presence of states constraints as the states are not preserved during reduction. If such constraints only involve a limited subset of states, the latter could be added to the outputs.

The examples show the versatility of the proposed framework. Concerning the problematic of gust loads alleviation, it enables to compare the efficiency of different actuators as well as the impact of their internal characteristics. It also enables to highlight the crucial impact of delay in that control problem. Such information could help to guide engineers in decision making at an early stage of novel aircraft design. Indeed, the latter is likely to rely heavily on advanced control functionalities to meet the ambitious environmental footprint reduction objectives. The interest of open-loop optimal control techniques is not restricted to this specific problem and we believe it could be more generally beneficial in the design process of various controlled systems.

The open-loop approach presented in this article leads to performances that could hardly be obtained in practice by feedback control. To explore further that direction, the convex parametrisation of stabilising controllers described e.g. in \cite[chap.7]{boyd:linear:1991} is of particular interest. However, it involves additional elements that are non-trivial to automatically derive such as the synthesis of an observer, the choice of a basis for the generating transfer matrix and a set of relevant feedback design objectives. These issues are currently under investigation.

\section*{Acknowledgements}
This work has been funded within the frame of the Joint Technology Initiative JTI Clean Sky 2, AIRFRAME Integrated Technology Demonstrator platform "AIRFRAME ITD" (contract N$^\circ$ CSJU-CS2-GAM-AIR-2014-15-01 Annex 1, Issue B04, October 2nd, 2015) being part of the Horizon 2020 research and Innovation framework programme of the European Commission.

\bibliographystyle{plain}

\bibliography{piblio.bib}   
\end{document}